\documentstyle[pre,twocolumn,aps,epsf,floats]{revtex} 

\begin{document}

\draft

\title{The Bak-Tang-Wiesenfeld sandpile model around the upper critical dimension}

\author{S. L\"ubeck\cite{SvenEmail} and K.~D. Usadel\cite{UsadelEmail} }
\address{
Theoretische Tieftemperaturphysik, 
Gerhard-Mercator-Universit\"at Duisburg,\\ 
Lotharstr. 1, 47048 Duisburg, Germany \\}
\author{\vskip -2\baselineskip\small(Received 19 June 1997)\break}
\author{\parbox{14cm}{\small
        \quad We consider the Bak-Tang-Wiesenfeld sandpile model
on square lattices in different dimensions ($D\le 6$).
A finite size scaling analysis of the avalanche
probability distributions yields the values
of the distribution exponents, the dynamical exponent,
and the dimension of the avalanches.
Above the upper critical dimension $D_u=4$
the exponents equal the known
mean field values.
An analysis of the area probability distributions
indicates that the avalanches are fractal 
above the critical dimension.
\hfill\break
\leftline{PACS number: 05.40.+j}}}
\address{\vskip +1.5\baselineskip}

%
%
%

\maketitle

\setcounter{page}{1}
\markright{\rm
Phys. Rev.~E 
, (to be published).
[accepted for publication]  }
\thispagestyle{myheadings}
\pagestyle{myheadings}

\section{Introduction}

Bak, Tang and Wiesenfeld \cite{BAK} introduced 
the concept of self-organized criticality (SOC)
and realized it with the so-called 'sandpile model' (BTW model).   
The steady state dynamics of the system is characterized by
the probability distributions
for the occurrence of relaxation clusters of a certain  size, area,  
duration, etc.
In the critical steady state these probability distributions
exhibit power-law behavior. 
Much work has been done in the two dimensional case.
Dhar introduced the concept of `Abelian sandpile models'
which allows to calculate the static properties of the model
exactly \cite{DHAR_2}, e.g.~the height probabilities, height 
correlations, number of steady state configurations, 
etc \cite{DHAR_2,MAJUM_1,PRIEZ_1,IVASH_1}.
Recently, the exponents of the probability distribution
which describes the dynamical properties of the system
were determined numerically \cite{LUEB_2}.
On the other hand both mean field
solutions (see \cite{VERGELES} and references therein) and 
the solution on the Bethe lattice \cite{DHAR_3} are well established
and both yield identical values of the exponents.
The mean field approaches are based on the assumption
that above the upper critical dimension $D_u$ the 
avalanches do not form loops and the avalanches propagation
can be described as a branching process \cite{ZAPPERI}.
Despite various theoretical and numerical efforts
the value of $D_u$ is still controversial.
In an early work, Obukhov predicted $D_u=4$ using an 
$\epsilon$-expansion renormalization group 
scheme \cite{OBUKHOV}.
Later D\'{\i}az-Guilera performed a momentum space analysis
of the corresponding Langevin equations which confirmed
$D_u=4$ \cite{DIAZ_1}.
Grassberger and Manna concluded from 
numerical investigations of the BTW model 
in $D\le 5$ the same result \cite{GRASS}.
In contrast, comparable simulations and 
the similarity to percolation
led several authors to the conjecture
that $D_u=6$ \cite{CHRIS_2} comparable to the related
forest fire model of Drossel and Schwabl (see \cite{CLAR_2}
for an overview).

In the present work we consider the BTW model
in various dimensions ($D\le 6$)
on lattice sizes which are significant
larger than those considered in previous
works \cite{GRASS,CHRIS_2,BENHUR}.
A finite size scaling analysis allows us to determine
the avalanche exponents, the dynamical exponent and 
to analyse whether 
the avalanche clusters are fractal.
Our analysis reveals that the upper critical
dimension is $D_u=4$ and
that the avalanches 
display a fractal behavior above $D_u$. 
We discuss the dimensional dependence
of the exponents and derive scaling relations.
Finally we briefly report results
of similar investigations of the
$D$-state model which is a possible generalization
of the two-state model introduced by Manna 
in two-dimensions \cite{MANNA_2}.
It is known that the BTW model and Manna's model
belong to different universality classes 
in $D=2$ \cite{BENHUR,LUEB_2}.

\section{Model and Simulations}

We consider the $D$-dimensional BTW model on a square lattice
of linear size $L$ in which integer variables $h_{\bf r}\ge 0$ represent
local heights.
One perturbes the system by adding particles at a randomly chosen site
$h_{\bf r}$ according to
\begin{equation}
h_{\bf r} \, \mapsto \, h_{\bf r}+1\, , \hspace{0.6cm} 
\mbox{with random }{\bf r}.
\label{eq:perturbation}
\end{equation}
A site is called unstable if the corresponding  height $h_{\bf r}$ 
exceeds a critical value $h_c$, i.e., if $h_{\bf r} \ge h_c$, 
where $h_c$ is given by $h_c=2D$.
An unstable site relaxes, its value is decreased by $h_c$
and the $2D$ next neighboring sites are increased by one unit, i.e.,
\begin{equation}
h_{\bf r}\;\to\;h_{\bf r}\,-\,h_c
\label{eq:relaxation_1}
\end{equation}
\begin{equation}
h_{nn,\bf r}\;\to\;h_{nn,\bf r}\;+\;1.
\label{eq:relaxation_2}
\end{equation}
In this way the neighboring sites may be activated and 
an avalanche of relaxation events may take place.
The sites are updated in parallel until all
sites are stable.
Then the next particle is added [Eq.~(\ref{eq:perturbation})].
We assume open boundary conditions with heights at the
boundary fixed to zero.

System sizes $L\le256$ for $D=3$, $L\le80$ for $D=4$,
$L\le36$ for $D=5$, and $L\le18$ for $D=6$
are investigated.
Starting with a lattice of randomly distributed heights 
$h\in\{0,1,2,...,h_c-1\}$ the system is perturbed according to
Eq.~(\ref{eq:perturbation})
and Dhar's 'burning algorithm' is applied in order to check if the 
system has reached the critical steady state \cite{DHAR_2}.
Then we start the actual measurements
which are averaged over
at least $2\times 10^6$ non-zero avalanches. 
We studied four different properties characterizing
an avalanche:  
the number of relaxation events $s$, 
the number of distinct toppled lattice site $s_d$ (area), 
the duration $t$, and the
radius $r$.
For a detailed description 
see \cite{LUEB_2} and references therein. 
In the critical steady state the corresponding probability 
distributions should obey power-law behavior 
characterized by exponents $\tau_s$, $\tau_d$,  $\tau_t$ and $\tau_r$
according to
\begin{equation}
P_s(s) \, \sim \, s^{-{\tau_s}},
\label{eq:prob_size}
\end{equation}
\begin{equation}
P_d(s_d) \, \sim \, {s_d}^{-{\tau_d}},
\label{eq:prob_distinct}
\end{equation}
\begin{equation}
P_t(t) \, \sim \, t^{-{\tau_t}},
\label{eq:prob_duration}
\end{equation}
\begin{equation}
P_r(r) \, \sim \, r^{-{\tau_r}}.
\label{eq:prob_radius}
\end{equation}
Because a particular lattice site may topple several times the 
number of toppling events exceeds the number of
distinct toppled lattice sites, i.e., $s \ge s_d$.
We will see that these multiple toppling events
can be neglected for $D\ge 3$ and 
the distribution $P_s(s)$ and $P_d(s_d)$
display the same scaling behavior.

Scaling relations for the exponents $\tau_s, \tau_d, \tau_t$ and
$\tau_r$ can be obtained if one assumes that the 
size, area, duration and radius scale as a 
power of each other, for instance 
\begin{equation}
t \, \sim \, r^{\gamma_{tr}}.
\label{eq:gam_tr}
\end{equation}
The relation $P_t(t) \mbox{d}t=P_r(r) \mbox{d}r$ for the
corresponding distribution functions then leads to
the scaling relation
\begin{equation}
{\gamma_{tr}}\;=\;\frac{\tau_r-1}{\tau_t-1}.
\label{eq:gam_tau_tr}
\end{equation}
The exponents $\gamma_{dr}$, $\gamma_{rs}$,
$\gamma_{sd}$ etc are defined in the same way.
The exponent $\gamma_{tr}$ is usually identified
with the dynamical exponent $z$ and various
theoretical efforts have been performed
to determine $z$ \cite{MAJUM_1,ZHANG_1,DIAZ_1}.
D\'{\i}az-Guilera \cite{DIAZ_1} concluded
from a momentum-space analysis of the corresponding
Langevin equations that the dynamical exponent of the 
BTW model is given by 
\begin{equation}
z \; = \; \frac{D\,+\,2}{3},
\label{eq:btw_z}
\end{equation}
which was already suggested by Zhang \cite{ZHANG_1}.
Numerical investigations suggest that Eq.~(\ref{eq:btw_z})
is valid \cite{BENHUR,LUEB_2}.
On the other hand Majumdar and Dhar \cite{MAJUM_1} used the
equivalence between the sandpile model
and the $q \to 0$ limit of the Potts model
to estimate $z=\frac{5}{4}$ in $D=2$ which
contradicts Eq.~(\ref{eq:btw_z}).

Christensen and Olami showed that inside an avalanche 
no holes can occur in the steady state \cite{CHRIS_2}
where a hole is a set of untoppled sites which are completely
enclosed by toppled lattice sites.
This implies for $D=2$ that the avalanches are
simply connected and compact.
For $D>2$ holes are still forbidden in the steady state 
but loops of toppled sites can occur.
Then the avalanches are no more simply connected (see below).
Even though no holes inside an 
avalanche cluster can occur 
it was already assumed that above the critical dimension $D_u$ 
the avalanches have the fractal dimension $4$ \cite{DHAR_3}.
Here, the propagation of an avalanche can not be considered
as a connected activation front of toppled sites.
The behavior is similar to an
branching process where 
disconnected arms propagate without forming loops. 
If the avalanche clusters are not fractal 
the scaling exponent $\gamma_{dr}$ which describes
how the number of toppled sites $s_d$ 
scales with the radius $r$ 
equals the dimension $D$.
Thus, the dimensional dependence of the exponent $\gamma_{dr}$ 
is an appropriate tool to investigate the developing fractal
behavior with increasing dimension.

The measurement of the 
probability distributions and the corresponding 
exponents
[Eq.~(\ref{eq:prob_size}-\ref{eq:prob_radius})]
is affected by the finite systems size.
For instance, the two dimensional BTW model displays a logarithmic
system size dependence of the distribution 
exponents \cite{MANNA_1,LUEB_2}.
Another example is the related two dimensional Zhang model \cite{ZHANG_1}
where the exponents depend on the inverse system size, i.e., the
corrections are of the relative magnitude of the
boundary $L^{-1}$ \cite{LUEB_3}.
In these cases the exponent of the infinite system
could be obtained by an extrapolation to the
infinite system size. 
If the values of the avalanche exponents $\tau$ are 
not affected by the finite system size the powerful 
method of finite size scaling would be applicable.
Here, the probability distributions 
[Eq.~(\ref{eq:prob_size}-\ref{eq:prob_radius})]
obey the scaling equation
\begin{equation}
P_x(x,L) \; = \;L^{-\beta_x} \, g_x (L^{-\nu_x}\,x), 
\label{eq:fss}
\end{equation}
with $x \in \{s,d,t,r\}$ and where $g_x$ is called 
the universal function \cite{KADANOFF}. 
The exponent $\tau_x$ is related to the scaling exponents
$\beta_x$ and $\nu_x$ via 
\begin{equation}
\beta_x=\tau_x \nu_x.
\label{eq:scaling_exponents}
\end{equation}
The exponent $\nu_x$ determines the cut-off behavior of the
probability distribution. 
If finite size scaling works all distributions $P_x(x,L)$
for various system sizes have to collapse,
including their cut-offs. 
Then the argument of the universal function $g$ has to
be constant, i.e., $x_{max} L^{-\nu_x}=\mbox{const}$.
Using the corresponding scaling relation [Eq.~(\ref{eq:gam_tau_tr})] 
yields $r_{max}^{\gamma_{xr}} L^{-\nu_x}=\mbox{const}$.
The cut-off radius $r_{max}$ should scale with the system size $L$
and finally one gets 
\begin{equation}
\nu_x \; = \; \gamma_{xr}.
\label{eq:gamma_nu}
\end{equation}
The advantage of the finite size scaling analysis is that it 
yields additionally to the avalanche
exponents $\tau_x$ the important scaling exponents 
$\gamma_{dr}$ and $\gamma_{tr}=z$.

\section{$D=3$}

\begin{figure}[h]
 \epsfxsize=8.6cm
 \epsfysize=7.0cm
 \epsffile{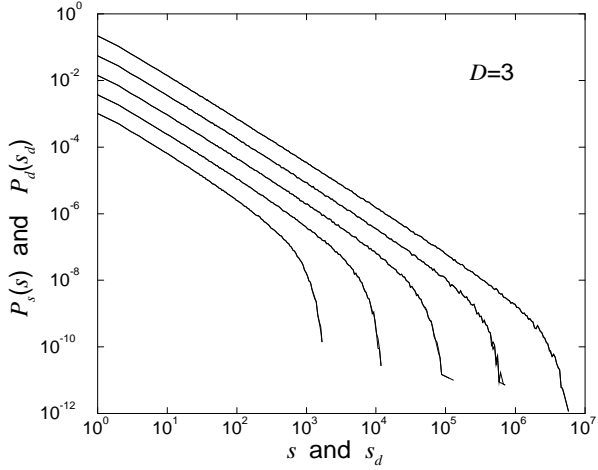}
 \caption{The probability distributions $P_s(s)$ and $P_d(s_d)$
          for $L\in\{16, 32, 64, 128, 256\}$. For $L<256$ the 
	  curves are shifted in the downward direction.
	  Note that there is 
	  no significant difference between both distributions,
	  i.e., multiple toppling events can be neglected in $D=3$.
 \label{p_s_p_d_d}} 
\end{figure}

In $D=3$ multiple toppling events, i.e., $s>s_d$,
occurs for less than 5\% of all avalanches
(nearly 42\% in $D=2$ and less than $0.1$\% in $D=4$).
These multiple toppling avalanches do not affect
the scaling behavior of the probability distribution
$P_s(s)$, in the sense that there is no significant 
difference between $P_s(s)$ and $P_d(s_d)$ 
(see Fig.~\ref{p_s_p_d_d}).
Thus one concludes that $\tau_d=\tau_s$ which
is confirmed by Ben-Hur and Biham who reported
that $\gamma_{sd}=1$ \cite{BENHUR}.

\begin{figure}[h]
 \epsfxsize=8.6cm
 \epsfysize=7.0cm
 \epsffile{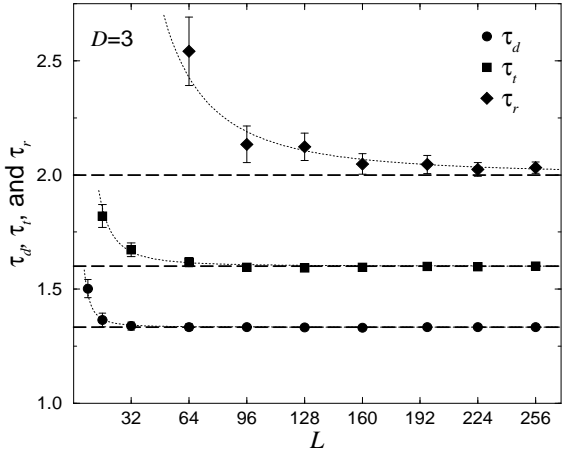}
 \caption{The system size dependence of the avalanches exponents
	  $\tau_d$, $\tau_t$, and $\tau_r$ for $D=3$.
	  The horizontal dashed lines correspond to the values $\frac{4}{3}$,
	  $\frac{8}{5}$, and $2$.
          The dotted lines are fits according to the equations 
          $\tau_x(L)=\tau_{x}-\mbox{const}L^{-2}$. 
 \label{tau_reg_3d}} 
\end{figure}

The exponents $\tau_d$, $\tau_t$, and $\tau_r$,
obtained from a power-law fit of the straight portion 
of the probability distributions 
[Eq.~\ref{eq:prob_distinct}-\ref{eq:prob_radius}],  
are plotted in Fig.~\ref{tau_reg_3d} for various system
sizes $L$.
The system size dependence vanishes quickly with
increasing $L$.
The dotted lines in Fig.~\ref{tau_reg_3d} corresponds to 
a $L^{-2}$ dependence of the avalanche exponents.
The finite size corrections are of the magnitude of the 
boundary term in three dimensions.
For $L\ge64$ the system size dependence of $\tau_d$ and $\tau_t$
is smaller than the statistical error of the
determination and the average of the exponents for $L\ge 64$
would be a good estimate of the values of the
infinite system. 
We obtain the values $\tau_d=1.333\pm0.007$
and $\tau_t=1.597\pm0.012$.
The value of $\tau_d$ is in agreement with previous investigations
based on smaller system sizes \cite{GRASS,BENHUR}.
The exponent $\tau_r$ seems to converge
in the vicinity of $2$ but the accuracy of this
measurement is not sufficient to decide whether
the value is exactly two. 
However, the following analysis lead us to the 
conclusion that $\tau_r=2$.

\begin{figure}[t]
 \epsfxsize=8.6cm
 \epsfysize=7.0cm
 \epsffile{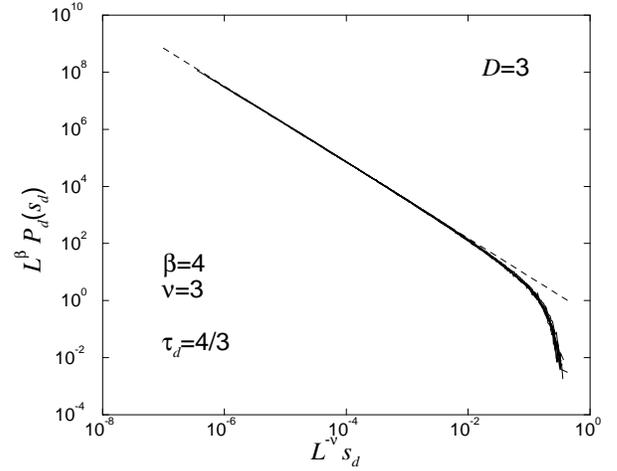}
 \caption{The scaling plot of the probability distribution $P_d(s_d)$
	  for $L\in\{64, 96, 128,..., 256\}$ and $D=3$. 
	  The dashed line corresponds to a power-law with an
	  exponent $\frac{4}{3}$.
 \label{tau_d_3d}} 
\end{figure}
\begin{figure}
 \epsfxsize=8.6cm
 \epsfysize=7.0cm
 \epsffile{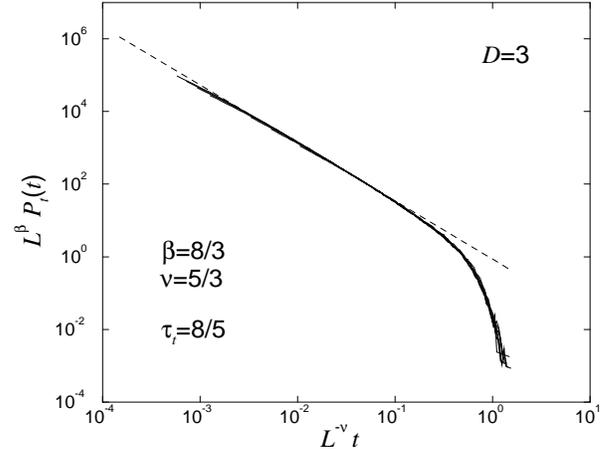}
 \caption{The scaling plot of the probability distribution $P_t(t)$
	  for $L\in\{64, 96, 128,..., 256\}$ and $D=3$. 
	  The dashed line corresponds to a power-law with an
	  exponent $\frac{8}{5}$.
 \label{tau_t_3d}} 
\end{figure}

Since the avalanche exponents $\tau_d$ and
$\tau_t$ display no significant
system size dependence for $L\ge 64$ the above mentioned finite-size
scaling analysis is applicable [Eq.~(\ref{eq:fss})].
The scaling plots of the distributions $P_d(s_d)$ and
$P_t(t)$ are shown in Fig.~\ref{tau_d_3d} and 
Fig.~\ref{tau_t_3d}.
One obtains a convincing data collapse of the various
curves corresponding to the different system sizes
for $\beta_d=4$, $\nu_d=3$, and $\beta_t=\frac{8}{3}$,
$\nu_t=z=\frac{5}{3}$, respectively.
Using Eq.~(\ref{eq:scaling_exponents}) the avalanches
exponents are given by $\tau_d=\frac{4}{3}$ and 
$\tau_t=\frac{8}{5}$.
These values are in agreement with our results
obtained from a direct determination of the 
exponents via regression.
The value $z=\frac{5}{3}$ agrees with Eq.~(\ref{eq:btw_z})
and $\nu_d=3$ reflects the fact that the
avalanches are not fractal.
This does not mean that the avalanche clusters
are still simply connected since
the avalanches can form loops.
But these rare loops do not contribute to the
scaling behavior.
Both scaling relations, $(\tau_r-1)=z(\tau_t-1)$ and
$(\tau_r-1)=\gamma_{dr}(\tau_d-1)$, confirm our
assumption that $\tau_r=2$.
In summary our direct measurements as well as the finite size scaling
analysis both yield that the avalanche exponents of the three dimensional
BTW model are consistent with the values $\tau_d=\tau_s=\frac{4}{3}$,
$\tau_t=\frac{8}{5}$, $\tau_r=2$, $z=\frac{5}{3}$, and
$\gamma_{dr}=3$.
All scaling relations which connect these exponents are 
fulfilled.

\section{$D\ge4$}

\begin{figure}[t]
 \epsfxsize=8.6cm
 \epsfysize=7.0cm
 \epsffile{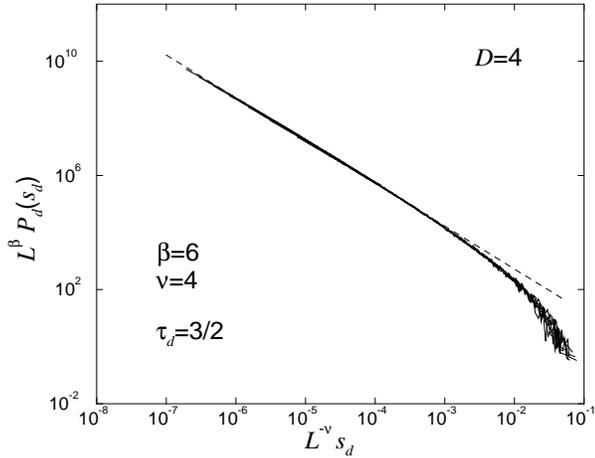}
 \caption{The scaling plot of the probability distribution $P_d(s_d)$
	  for $L\in\{16,24,32,48,...,80\}$ and $D=4$. 
	  The dashed line corresponds to a power-law with the 
	  exponent $\frac{3}{2}$.
 \label{tau_d_4d}} 
\end{figure}

\begin{figure}[t]
 \epsfxsize=8.6cm
 \epsfysize=7.0cm
 \epsffile{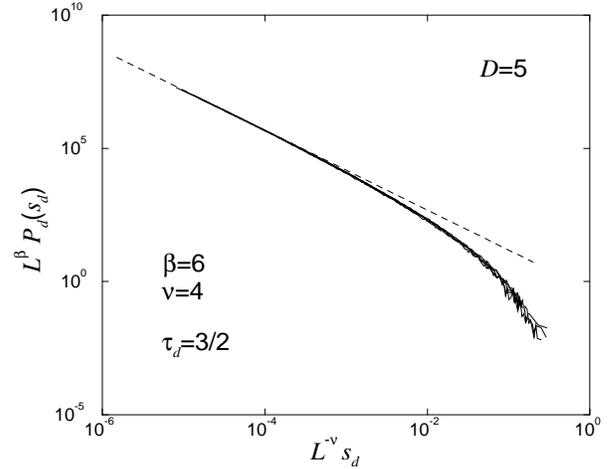}
 \caption{The scaling plot of the probability distribution $P_d(s_d)$
	  for $L\in\{8,16,20,24,...,36\}$ and $D=5$. 
	  The dashed line corresponds to a power-law with the 
	  exponent $\frac{3}{2}$.
 \label{tau_d_5d}} 
\end{figure}

\begin{figure}
 \epsfxsize=8.6cm
 \epsfysize=7.0cm
 \epsffile{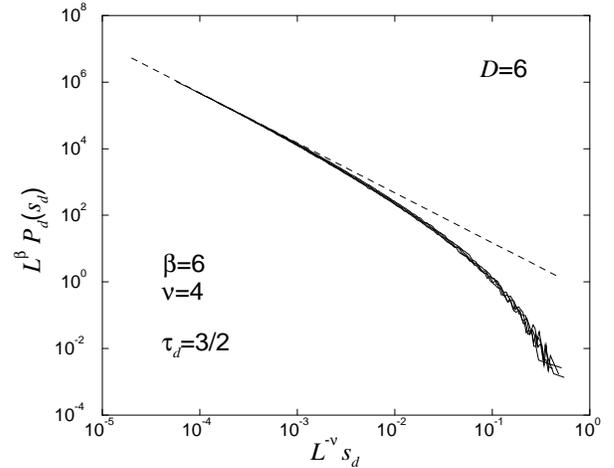}
 \caption{The scaling plot of the probability distribution $P_d(s_d)$
	  for $L\in\{8,10,12,...,18\}$ and $D=6$. 
	  The dashed line corresponds to a power-law with the 
	  exponent $\frac{3}{2}$.
 \label{tau_d_6d}} 
\end{figure}

Focusing our attention to the area and duration
probability distribution we find that 
finite size scaling works quite well again.
In Fig.~\ref{tau_d_4d}, \ref{tau_d_5d}, and \ref{tau_d_6d} we present
the scaling plots of the avalanche distribution
$P_d(s_d)$ for $D=4$, $D=5$, and $D=6$.
In all cases one gets a satisfying data collapse for 
$\beta_d=6$ and $\nu_d=4$, i.e., the corresponding 
avalanches exponent equals the mean field 
value $\tau_d=\frac{3}{2}$.
A similar analysis displays that the scaling exponents of the
duration distribution $P_t(t)$ are given by $\beta_t=4$ and $\nu_t=2$
resulting in $\tau_t=2$ (not shown).
The avalanche exponents of the BTW model
in $D\ge4$ agree with the mean field exponents 
$\tau_d=\frac{3}{2}$, $\tau_t=2$, $z=2$, 
and the upper critical dimension is $D_u=4$.
All exponents are listed in 
Table~\ref{TABLE_BTW}.
An analysis of the probability distribution
$P_r(r)$ and the determination of the exponent 
$\tau_r$ remains outside the scope of this paper
because the considered system sizes (limited by 
computer power) are too small.
For instance, in the case of $D=4$ the largest
considered system sizes is $L=80$.
The corresponding distribution $P_r(r)$ exhibits
a very small power-law region (less than a half
decade), forbidding
any accurate determination of $\tau_r$.


\begin{figure}[t]
 \begin{minipage}[t]{17.8cm}
 \epsfxsize=16.0cm
 \epsfysize=16.0cm
 \epsffile{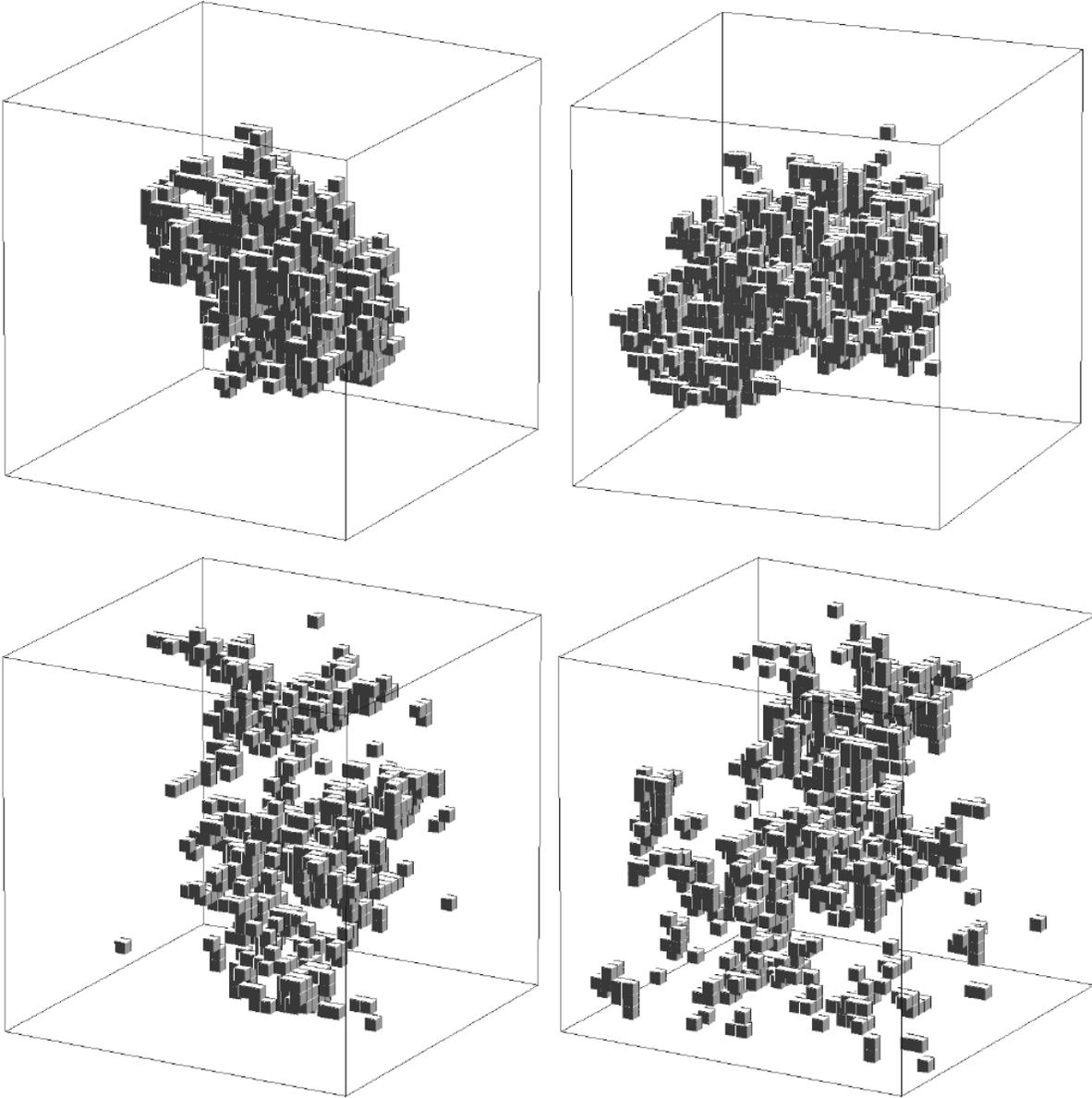}
 \caption{Snapshots of four arbritrary chosen avalanche clusters
	  for $D=3$ (upper left), $D=4$ (upper right), and $D=5$
	  (lower left and right).
	  For $D\ge 4$ three dimensional cuts through the center 
	  of mass are shown. 
          In the three dimensional avalanche a loop can be seen
	  in the upper left part of the avalanche cluster. 
 \label{btw_aval}}  
\end{minipage}
\end{figure}

The value $\gamma_{dr}=4$ corresponds to the fact that
the avalanches of the BTW model display a fractal behavior 
above the critical dimension $D_u$,
whereby the area scales with the radius according to
$s_d \sim r^4$, independently of the embedding dimension $D$.
For $D\le D_u$ the avalanches are not fractal. 
We display this developing fractal behavior in 
Fig.~\ref{btw_aval}, where four arbritrary chosen 
avalanche clusters are shown for three different dimensions.
For $D\ge4$ we plotted three dimensional cuts through
the center of mass of the avalanche clusters.
The isolated islands which appear in the avalanche snapshots 
for $D\ge 4$ are caused by the three dimensional cuts.
In all cases the system size is $L=32$
and the area of the plotted avalanches
is $s_d=1520$ in $D=3$,
$s_d=17500$ in $D=4$,
and $s_d=201000$ in $D=5$,
i.e., $s_d^{1/D}$ is nearly fixed.
If the avalanches are not fractal in all dimensions
the scaling relation $s_d\sim r^d$ holds for all $D$ and
the radius should be independent of the embedding dimension.
One can see 
see from Fig.~\ref{btw_aval} that the radius
of the shown avalanches is roughly the same for $D=3$ and $D=4$.
Despite some loops (e.g.~in the upper left part 
of the plotted three dimensional avalanche)
the avalanche clusters look nearly compact.
In the five dimensional case the clusters display a
fractal behavior.
The radius seems to be larger compared to the lower dimensional
cases indicating that the equation $s_d\sim r^D$ 
does not hold in $D=5$.
Of course these snapshots only illustrate the developing 
fractal behavior.\\ 

\vspace{2cm}
$ \quad $ \\

\vspace{17cm}
Our results are in contrast to previous investigations
performed by J\'{a}nosi and Czir\'{o}k \cite{JANOSHI_2}.
They calculated the number of toppled site $N(r)$ 
inside a sphere with radius $r$.
The sphere is centered at the center of mass of the
avalanche cluster.
The fractal dimension $D_f$ is obtained from the scaling 
law $N(r) \sim r^{D_f}$.
Considering one system size ($L=100$ in $D=3$) 
they found that the fractal dimension is given
by $D_f\approx 2.75$, i.e., the avalanches already display
a fractal behavior in three dimensions.
We performed the same analysis and reproduced their
results within the error-bars.
Analyzing various system sizes, however, 
we find that the apparent fractal dimension
depends on the system size
and tends to $D_f=3$ with increasing $L$ (not shown)
in agreement with our results, discussed above.

\section{discussion}

\begin{table}[t]
\caption{Values of the exponents 
of the BTW model in various dimensions.}
\label{TABLE_BTW}
\begin{tabular}{cccccc}
               & $D=2$          & $D=3$         & $D=4$         & $D=5$         & $D=6$         \\  
\tableline \\
$\tau_{s}$     & 1.293          & $\tau_{d}$    & $\tau_{d}$    & $\tau_{d}$    & $\tau_{d}$    \\ \\
$\tau_{d}$     & $\frac{4}{3}$  & $\frac{4}{3}$ & $\frac{3}{2}$ & $\frac{3}{2}$ & $\frac{3}{2}$ \\ \\
$\tau_{t}$     & $\frac{3}{2}$  & $\frac{8}{5}$ & $2$           & $2$           & $2$           \\ \\
$\tau_{r}$     & $\frac{5}{3}$  & $2$           &               &               &               \\ \\
$z$            & $\frac{4}{3}$  & $\frac{5}{3}$ & $2$           & $2$           & $2$           \\ \\
$\gamma_{dr}$  & $2$  		& $3$           & $4$           & $4$           & $4$           \\ 
\end{tabular}
\end{table}

\begin{figure}
 \epsfxsize=8.6cm
 \epsfysize=7.0cm
 \epsffile{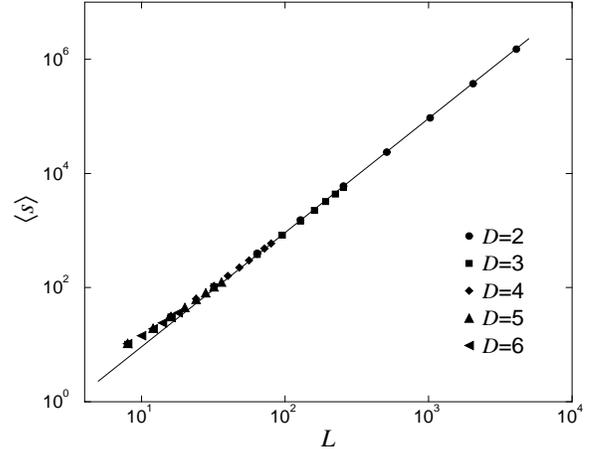}
 \caption{The average avalanche size $\langle s \rangle_L$ as a
	  function of the system size in various dimensions.
	  The solid line corresponds to the power-law
	  $\langle s \rangle_L \sim L^2$ which is
	  exactly in $D=2$ \protect\cite{DHAR_2}.
 \label{mean_s_L_dim}} 
\end{figure}

In the following we examine the avalanche exponents
as a function of the dimension $D$.
Consider the average avalanche size 
\begin{equation}
\langle s \rangle_L \; = \;
\int s \, P_s(s,L) \, \mbox{d}s.
\label{eq:aver_size}
\end{equation}
Using the finite size scaling ansatz [Eq.~(\ref{eq:fss})]
which works for $D\ge 3$ one gets \cite{KADANOFF}
\begin{equation}
\langle s \rangle_L \; \sim \; L^{2 \nu_s - \beta_s}
\; = \; L^{\gamma_{sr}(2-\tau_s)},
\label{eq:aver_size_exp}
\end{equation}
if $\tau_s<2$.
On the other hand it is known exactly \cite{DHAR_2} that 
$\langle s \rangle_L \sim L^2$ in $D=2$ and
arguing that in undirected models particles 
diffuses out to the boundary one gets the same
result independent of the dimension \cite{KADANOFF}.
Like Grassberger and Manna \cite{GRASS} we plot
in Fig.~\ref{mean_s_L_dim} the average avalanches size as a function of the 
system size for various dimensions.
Except of deviations for small system sizes 
all data points collapse on a single curve.
Thus one concludes that the equation $2=\gamma_{sr}(2-\tau_s)$ 
is fulfilled.
Neglecting multiple toppling ($\tau_s=\tau_d$ and 
$\gamma_{sr}=\gamma_{dr}$) which is valid for
$D\ge 3$ and using that the avalanches are not fractal
($\gamma_{dr}=D$) which is fulfilled for $D\le D_u$ one gets
\begin{equation}
\tau_d \; = \; 2 \, -\, \frac{2}{D},
\label{eq:tau_d_dim}
\end{equation}
for $3 \le D\le D_u$ \cite{BEM_1}.
This equation was already derived in the 
continuum limit by Zhang using energy
conservation and the local nature of energy transfer \cite{ZHANG_1}.
Now we see that the failure of this equation for $D=2$ is 
caused by multiple toppling events which are essential 
in the two dimensional model only.
For $D\ge 3$ multiple toppling can be neglected and 
Eq.~(\ref{eq:tau_d_dim}) is fulfilled.
Using 
\begin{equation}
z\, (\tau_t-1) \; = \; \gamma_{dr} \, (\tau_d-1)
\label{eq:tau_d_t}
\end{equation}
and Eq.~(\ref{eq:btw_z})
the duration exponent $\tau_t$ is given by
\begin{equation}
\tau_t \; =\; 4\,\frac{D-1}{D+2},
\label{eq:tau_t_dim}
\end{equation}
again for $3 \le D\le D_u$.

\begin{table}[b]
\caption{Values of the exponents 
of $D$-state model in various dimensions.}
\label{TABLE_MANNA}
\begin{tabular}{ccccc}
               & $D=2$           & $D=3$           & $D=4$         & $D=5$         \\  
\tableline \\
$\tau_{s}$     & $\frac{14}{11}$ & $\approx\frac{14}{10}$  & $\tau_{d}$    & $\tau_{d}$    \\ \\
$\tau_{d}$     & $\frac{11}{8}$  & $\approx\frac{13}{9}$  & $\frac{3}{2}$ & $\frac{3}{2}$ \\ \\
$\tau_{t}$     & $\frac{3}{2}$   & $\approx\frac{37}{21}$ & $2$           & $2$           \\ \\
$\tau_{r}$     & $\frac{7}{4}$   & $\approx\frac{7}{3}$   &               &               \\ \\
$z$            & $\frac{3}{2}$   & $\approx\frac{7}{4}$   & $2$           & $2$           \\ \\
$\gamma_{dr}$  & $2$  		 & $\approx 3$            & $4$           & $4$           \\ 
\end{tabular}
\end{table}

Finally we briefly report results of similar
investigations of the related $D$-state sandpile model 
based on Manna's two-dimensional two-state model 
\cite{MANNA_2}.
Here the critical height $h_c$ equals the dimension $D$
and an unstable site relaxes to zero, whereby the particles
are distributed randomly among the nearest neighbors.
Again we find that the upper critical dimension is $D_u=4$.
In contrast to the BTW model the dimensional dependence
of the dynamical exponent is given by $z=(D+4)/4$.
Our preliminary results for $D=3$ are
$\tau_s \approx \frac{14}{10}$,
$\tau_d \approx \frac{13}{9}$, $\tau_t \approx \frac{37}{21}$,
$\tau_r \approx \frac{7}{3}$, and $\gamma_{dr}\approx 3$.
We find that $\tau_d$ is definitely larger than 
$\tau_s$ (in agreement with \cite{BENHUR}), i.e.,
multiple toppling events are relevant in the three dimensional
model.
Because in the $D$-state model the toppling processes are isotropic 
on average only holes inside an avalanche cluster can occur.
But nevertheless, we find that $\gamma_{dr}=D$ for
$D\le D_u$, i.e., these holes occur only on finite
sizes and do not contribute to the scaling behavior.
Above the critical dimension $D_u=4$ the avalanches have
fractal dimension 4.
In $D=4$ and $D=5$ the model is characterized by the 
mean field exponents, comparable to the BTW model.
The values of the exponents are listed in 
Table~\ref{TABLE_MANNA}.

\section{Conclusions}

We studied numerically the dynamical properties of the BTW model on a 
square lattice in various dimensions.
Using a finite size scaling analysis we determined the
probability distribution exponents, the dynamical exponent, and the 
dimension of the avalanches.
Our analysis reveals that multiple toppling events
are relevant in the low dimensional case only
and can be neglect for $D\ge 3$.
For $D=3$ the exponents are given by $\tau_r=2$,
$\tau_t=\frac{8}{5}$, $\tau_d=\frac{4}{3}$,
and $z=\frac{5}{3}$.
For $D\ge4$ the exponents agree with the 
mean field and Bethe lattice exponents, respectively.
We conclude from our numerical results that below
the critical dimension the 
dynamical exponent $z$ is given by $z=(D+2)/3$.
The avalanche clusters are simply connected for $D=2$ only.
For $D > 2$ loops occur but do not contribute
to the scaling behavior until the embedding dimension
exceeds the upper critical dimension $D_u$.
Above $D_u$ the avalanches are fractal
with the fractal dimension $4$.

\acknowledgments
This work was supported by the
Deutsche Forschungsgemeinschaft through
Sonderforschungsbereich 166, Germany.


\begin{references}
\bibitem{BEM_1}	    Grassberger and Manna performed a similar
		    analysis \protect\cite{GRASS}.
		    Instead of the finite size scaling equation 
	            they calculated the average system size via the
		    probability distribution [Eq.~(\protect\ref{eq:prob_distinct})].
		    Therefore, they had to estimate the upper limit
		    of integration ($s_d \le L^d$) and got the
		    inequality $\tau_d \le 2 - 2/d$ only.
\end{references}
\end{document}